\documentclass[onecolumn,showpacs,preprintnumbers,amsmath,amssymb]{revtex4}

\usepackage{graphicx}
\usepackage{dcolumn}
\usepackage{bm}

\def\be{\begin{equation}}
\def\ee{\end{equation}}

\begin{document}

\title{The Fuzzy Sphere: Star Product Induced From Generalized Squeezed States.}

\author{Musongela Lubo}
 \affiliation{The Abdus Salam International Centre for Theoretical Physics I.C.T.P\\
	P.O.Box
586\\
34100 Trieste, Italy.\\
}
\email{muso@ictp.trieste.it}

\date{\today}
\begin{abstract}
A family of states built from the uncertainty principle
on the fuzzy sphere has been shown
to reproduce the stereographic projection in the large $j$ limit. These
generalized squeezed states are used to construct an associative star 
product which involves a finite number of derivatives on its primary functional space. It is written in terms of a variable on the complex plane.
We show that it actually coincides with the one found by
H.Gross and P.Presnajder in the simplest cases, endowing the later
with a supplementary physical interpretation. We also show how the spherical harmonics
emerge in this setting.
\end{abstract}

\maketitle

\section{Introduction.}

The study of non commutative structures in Physics was initially motivated by the idea that they 
may provide a geometric regularization of the divergences encountered in field theory. This has 
been fulfilled in some models, the most prominent being the fuzzy sphere \cite{prea,preb}. The interest in 
this field of theoretical physics has been recently boosted by the realization that such spaces 
emerge naturally in string theory and matrix models \cite{seiberg}. These spaces also provide models in which Lorentz 
symmetry is true only  at low energies. But, they possess a robust 
mathematical structure which makes them more elegant and more predictive than  ad hoc models.

Contacts with experiments and observations are difficult for two major reasons. The first one is 
that the mathematical difficulties encountered when studying such spaces has essentially led to
restrict the attention to toy models. The case of the 
non commutative flat Minkowski space is  of course the most important exception, thanks to the
fact that one can then make a rotation which keeps all the fuzziness in one plane. 
It was recently proposed that modified dispersion relations similar to the ones inferred from the 
fuzzy sphere could be of some help when tackling the problem posed by the gamma ray bursts 
\cite{sidha}. Although this idea has been proposed in some ad hoc models, the fact that a fuzzy structure
can have testable predictions is interesting. 

The topics related to non commutativity are numerous. For example, their stability when they are
obtained from a matrix model is the focus of many recent works like \cite{takeh} and so are 
the U.V/I.R mixing  \cite{castro}, the links with string theory \cite{biga}, the properties of their
solitons \cite{urs} and the behavior of exotic particles like anyons \cite{anyon}.

The star product is one of the key ingredients in non commutative theories \cite{stei}. It allows a simple
formulation of field theory on fuzzy structures. The most famous  is the one obtained by Moyal
\cite{moyal1,moyal2,voros} on a non commutative plane. When studying Q.F.T on this space, a simple 
prescription consists in
replacing all ordinary products by their star counterparts.

In this paper we derive a star product on the fuzzy sphere. This has been done by many authors
\cite{pre1,pre2,wata1,wata2,wata3,
bala1,bala2,bala3,ho,arik,star1,star2,star3,star4,star5,star6} and in various ways. Our approach is 
essentially   close to the one used by
G.Alexanian, A.Pinzul and A.Stern \cite{pinzul1} in the sense that it relies on the integral approach using generalized coherent
states. The first difference relies  in the states used. The basis of our construction is provided
by the generalized squeezed states found in \cite{lubo1}. They were obtained by relaxing the 
Heisenberg uncertainty and were shown to reproduce the stereographic projection in the large $j$ limit.
The second difference is the fact that we can show how the star product so built, although it
contains an infinite number of derivatives at first sight, 
this is not the case when it
acts on its primary functional space, which by definition contains the mean values of all the 
operators of the fuzzy space on the generalized coherent states. This  character 
resembles  the one displayed by  the star
product obtained in \cite{pre1}. Although our formulation relies on a single complex variable
rather than three dependent real ones, we will show that the two products are actually the same, at least
when non commutativity is strong.

This paper is organized as follows. The second section summarizes the properties of the generalized 
squeezed states which will be needed in this work. In the third section
we give a short reminder of the integral form of the star products built from generalized coherent 
states. The fourth section is devoted to the application of this formalism to the two dimensional
fuzzy sphere. We then derive a finite expression for the star product when it is restricted to its
natural set of definition. The sixth section makes the link between our approach
and the one followed by H.Grosse and P.Presnajder. In the seventh section we analyze the link between the primary functional
space of our star product and the spherical harmonics which are obvious tools when dealing with
a system displaying rotational symmetry. The possible extensions of our work and some discussions are
the subject of the eighth section.

\section{Generalized squeezed states on the fuzzy sphere.}

The fuzzy sphere is obtained by promoting the Cartesian coordinates to non commuting
operators satisfying the following  relations
 \cite{madore1,madore2}:
\begin{eqnarray}
\label{eqb1}
[ \hat x_k , \hat x_l ] &=& \frac{i R}{\sqrt{j(j+1)}} \epsilon_{klm} \hat x_m \quad , \quad
\delta^{l k} \hat{x}_l \hat{x}_k = R^2  \quad , \nonumber\\
& & {\rm with}
\quad j \quad  {\rm integer} \quad {\rm or } \quad {\rm half-integer} \quad
{\rm and} \quad k,l,m =1,2,3 \quad .
\end{eqnarray}
It is important that the symmetries of the classical and the fuzzy sphere are the same;
this means one essentially works with representations of the group $SU(2)$. The 
value
of the radius $R$ will be related to the only value permitted for the Casimir operator on the 
representation used ; we will chose  rescaled variables such that it can be set equal to
one.

The differences concerning the construction of coherent states on the non commutative plane
and the fuzzy sphere has been considered by many others, among which \cite{nekra}. While on the
first space the saturation of the Heisenberg uncertainty is enough to build a family of
states with the right properties, one has to   resort, on the second, to the weaker inequalities
\cite{lubo1}
\begin{eqnarray}
\label{eqb2}
  (\Delta x_1)^2  (\Delta x_2)^2  & \geq& \frac{1}{4}  
  \left[  (  \langle  \{ \hat{x}_1 , \hat{x}_2  \} \rangle
    - 2
\langle \hat{x}_1 \rangle  \langle \hat{x}_2 \rangle )^2 + 
\vert \langle [ \hat{x}_1 , \hat{x}_2] \rangle  \vert^2 
\right]  \quad ,  \nonumber\\
(\Delta x_2)^2  (\Delta x_3)^2  &\geq& \frac{1}{4}  
  \left[  (  \langle  \{ \hat{x}_2 , \hat{x}_3  \} \rangle
    - 2
\langle \hat{x}_2 \rangle  \langle \hat{x}_3 \rangle )^2 + 
\vert \langle [ \hat{x}_2 , \hat{x}_3] \rangle  \vert^2 
\right]   \quad , \nonumber\\
(\Delta x_3)^2  (\Delta x_1)^2  &\geq& \frac{1}{4}  
  \left[  (  \langle  \{ \hat{x}_3 , \hat{x}_1  \} \rangle
    - 2
\langle \hat{x}_3 \rangle  \langle \hat{x}_1 \rangle )^2 + 
\vert \langle [ \hat{x}_3 , \hat{x}_1] \rangle  \vert^2 
\right]  \quad . 
\end{eqnarray}

The generalized squeezed state takes the form
\begin{equation}
\label{eqb3}
\vert \zeta \rangle = \frac{1}{(2 j+ \zeta \bar\zeta)^j} \sum_{m=-j}^j  \, R_m \, \zeta^{-m+j}
\, \vert m \rangle \quad ,
\end{equation}
 where the real constants $R_m$ are given by
\begin{eqnarray}
\label{eqb4}
R_m &=&  (2 j)^{\frac{1}{2}(m+j)} 
       \left( \frac{\Gamma(2 j+1)}{\Gamma(j-m+1) \Gamma(j+m+1)} \right)^{1/2}
  \quad .
\end{eqnarray}

While the simultaneous saturation of the Heisenberg uncertainties on the fuzzy sphere leads
to the empty set, doing the same with the weaker relations displayed in Eq.(\ref{eqb2})
leads to a family of states parameterized by a complex number $\zeta$. This implies  a 
mapping from the complex plane to our family of states. Using the variable defined by $\zeta= \sqrt{2 j} \alpha$ gives to the mean values
$\langle \hat{x}_k \rangle$ the expression of the stereographic 
projection in the limit $j\rightarrow \infty$ . The uncertainties on the positions 
vanish in the same limit.

We record for future use the expression of the scalar product between two such generalized
squeezed states:
\begin{equation}
\label{eqb5}
\langle \zeta  \vert \eta \rangle = \frac{(2 j+ \bar\zeta \eta)^{2 j}}
{(2 j+ \bar\zeta \zeta)^{j} (2 j+ \bar\eta \eta)^{j}} \quad .
\end{equation}
Taking the  limit $j\rightarrow \infty$, one finds this scalar product tends to a Gaussian 
as on the non commutative plane.

It is obvious that our construction is not as general as the ones developed in \cite{klau,kon,za}. 

\section{Coherent states and the integral form of the star product}

\subsection{A reminder}

As stated in the introduction, the star product is one of the most important ingredients in non commutative field theory.
For example, theories on the non commutative plane can be obtained by taking the usual
action where all the usual products have been replaced by the Moyal product \cite{nekra}.
The Moyal star product is based on the Weyl correspondence which incorporates plane waves in a special
way.  

When working with coherent states, the star product is first defined for a special class 
of functions(which we shall call the {\bf primary functional space}). To any such 
function $f$ one assigns an operator $\hat W[f]$ and by definition \cite{stei}:
\begin{equation}
\label{eqc1}
\hat{W} [ {\cal A} \star {\cal B}] = \hat{W} [ {\cal A} ]  \hat{W} [ {\cal B} ]  \quad .
\end{equation}
The primary functional space will be obtained by taking the mean values of  quantum 
operators on the coherent states:
\begin{equation}
\label{eqc2}
 {\cal A} (\zeta,\bar\zeta) = \langle \zeta \vert  \hat{W} [{\cal A}] 
 \vert \zeta \rangle \quad , \quad
 \hat{W}^{-1} [ \hat{A} ] (\zeta, \bar\zeta) =  \langle \zeta \vert \hat A \vert 
 \zeta \rangle
 \quad .
\end{equation}
Using these two formulas, one obtains 
\begin{equation}
\label{eqc3}
({\cal A} \star {\cal B})(\zeta,\bar\zeta) = 
\langle \zeta \vert \hat W [{\cal A}] \hat W[{\cal B}] \vert \zeta 
\rangle \quad .
 \end{equation}  
 The 
associativity of the star product is then guaranteed by the associativity of the
 product of the quantum operators.
To study Q.F.T, one needs to express this  product in terms of 
differential operators acting on the functions $f_{\hat A},f_{\hat B}$. 
This can be done if one has a family of states $ \vert\eta \rangle$ to which 
is associated 
a decomposition
 of the unity operator.
This requires the knowledge of a weight:
\begin{equation}
\label{eqc4}
 \int d\mu(\eta, \bar\eta) \vert\eta \rangle \langle\eta \vert = 1 \quad .
\end{equation}
If $\vert\eta \rangle $ is, up to a normalization factor depending on the 
absolute value $ \vert\eta \vert$,
analytic in $\eta$, one is led to 
the following expression of the star product \cite{pinzul1}:
\begin{equation}
\label{eqc5}
{\cal A}(\zeta,\bar\zeta) \star {\cal B}(\zeta,\bar\zeta) = 
\int d\mu(\eta, \bar \eta)     \vert \langle\zeta \vert \eta \rangle  \vert^2 
\left( e^{-\zeta \partial_\eta} e^{ \eta \partial_\zeta} {\cal A}(\zeta,\bar \zeta) \right)
\left( e^{- \bar\zeta \partial_{\bar \eta}} e^{ \bar \eta \partial_{\bar\zeta}} 
{\cal B}(\zeta,\bar \zeta) \right) \quad .
\end{equation}

From this one sees that the derivatives with respect to the variable $\zeta$ will act only
on the first function ${\cal A}$ while those with respect to its conjugate $\bar\zeta$ will
act only on the second one i.e ${\cal B}$.

The last formula can then be used to extend the star product to functions out of the
functional space defined by Eq.(\ref{eqc2}). We will illustrate this point later. 

At this stage one important point is to be noticed. The formula displayed
in Eq.(\ref{eqc5}) contains an infinite number of derivatives. But, as we shall show, the
functional space defined by Eq.(\ref{eqc2}) is finite dimensional. This means that all the 
derivatives above a fixed order ($2 j$) can be written as combinations of derivatives below that
order. This is the reason why we will obtain a star product containing a finite number of
terms.
This fact is of great importance for field theory since it means that if one restricts
oneself to  the  primary set of functions, the field theory will contain a finite number
of derivatives, contrary to 
what happens with the Moyal star product on the non commutative plane.
 
The formula displayed in Eq.(\ref{eqc5}) does not seem to use the Wely map. However
this map is crucial in defining the star product in Eq.(\ref{eqc3}). We used here the
explicit form of the map given in Eq.(\ref{eqc2}). We will come back to this point
in our illustrations concerning the link with the 
Grosse Presnajder product.

\subsection{Application to the fuzzy sphere.}

A simple way of defining an extension of the star product  relies on a decomposition of the unity operator
 as pointed out in the third section.
We need a measure $d\mu(\zeta,\bar\zeta)$ satisfying Eq.(\ref{eqc4}).
This extension will  contain  an infinite number  of 
derivatives. We assume from now on that $j \geq 2$.

One of the key relations we will use often  stems from the decomposition
of the unity operator:
\begin{equation}
\label{eqd1}
 \langle n \vert m \rangle = \delta_{m,n}
 \Longrightarrow 
\int d\mu(\eta, \bar\eta) \frac{\eta^{(-n+j)} \bar\eta^{(-m+j)}}{(2 j+ \bar\eta \eta)^{2 j}} 
 = \frac{1}{R_n R_m}  \delta_{n,m} \quad , \quad {\rm for} \quad - j \leq m, n \leq j \, .
\end{equation}
At this stage, this is the only constraint on the measure. Of course, one can find more 
than one solution to this set of $(2 j+1) \times (2 j+1)$ equations. Each such solution will
lead to a different extension of the star product. Presumably, these star products will be
related. In fact, from any star product $\star$ and any differential operator
$D$ one can build 
another associative star product $\star^{\prime}$ in the following way:
\begin{eqnarray}
\label{eqd2}
f \star^{\prime} g = D(D^{-1} f \star D^{-1} g )  \quad ;
\end{eqnarray}
the two are said to be gauge equivalent \cite{kon}. Nevertheless, the restrictions of these star products
on the primary functional space must coincide since Eq.(\ref{eqc2}) does not involve the measure.

Integrating with respect to the variable $\eta$ (see Eq.(\ref{eqd1})), the star product can now 
be recast in the form
\begin{eqnarray}
\label{eqd5}
& & {\cal A}(\zeta,\bar\zeta) \star {\cal B}(\zeta,\bar\zeta)
=  {\cal A}(\zeta,\bar\zeta) \, {\cal B}(\zeta,\bar\zeta) +
{\cal A}(\zeta,\bar\zeta) \, 
\left( \sum_{l_1=1}^{\infty} 
\tilde{D}_{l_1} \partial_{\bar\zeta}^{l_1}  {\cal B}(\zeta,\bar\zeta)
\right)  \nonumber\\
&+& \left(  \sum_{k_1=1}^{\infty} 
\tilde{C}_{k_1} \partial_{\zeta}^{k_1} {\cal A}(\zeta,\bar\zeta) \right)  
 {\cal B}(\zeta,\bar\zeta)  
 +  \sum_{k_1=1}^{\infty}  \sum_{l_1=1}^{\infty}   
\tilde{M}_{k_1,l_1} \partial_{\zeta}^{k_1} {\cal A}(\zeta,\bar\zeta) 
 \partial_{\bar\zeta}^{l_1} {\cal B}(\zeta,\bar\zeta) \quad .
\end{eqnarray}

The coefficient functions are rational expressions of the variable $\zeta$ which 
parameterizes the generalized squeezed states:

We now use the Kronecker $\delta$  to  simplify these formulas. We  also extract the
quantity $\zeta \bar\zeta$ whenever possible. The first coefficient reads:
\begin{eqnarray}
\label{eqd9}
& & \tilde{C}_{k_1} = \frac{\zeta^{k_1}}{(2 j+ \zeta \bar\zeta)^{2 j}}
\sum_{m_1=0}^{2 j} \tilde{\gamma}_{k_1,m_1} (\zeta \bar\zeta)^{m_1} \quad , {\rm where} \nonumber\\
\tilde{\gamma}_{k_1,m_1} &=& \frac{(2 j)! (2 j)^{2 j-m_1}}{\Gamma(m_1+1) \Gamma{(2 j-m_1+1)}} 
 \left( \frac{1}{\Gamma(k_1+1)}  (1 - \Theta(k_1 - 2 j))   \right. \nonumber\\
& & \left. + (-1)^{k_1+m_1}
\sum_{m_2}    \frac{(-1)^{m_2}}{\Gamma{(k_1+m_1-m_2+1) \Gamma(m_2-m_1+1)}}   \right) \quad .
\end{eqnarray}
The last sum is performed on integers $m_2$ verifying the inequality
\begin{equation}
\label{eqd10}
1 \leq k_1 + m_1 - m_2 (\equiv k_2) \leq k_1 \quad 
\end{equation}
and  the  step function $\Theta$ appears due to the fact that if $k_1>2 j$, the Kronecker
symbol $\delta_{k_1+m_1,m_2}$ automatically vanishes, $m_2$ being smaller than $2 j$. The form
of the coefficient $C_k$ has a simple structure: it is the product of the $k^{\rm th}$ power of $\zeta$
with a polynomial of degree $2 j$ in the quantity $\zeta \bar\zeta$, the result being 
divided by the ubiquitous expression $(2 j+\zeta \bar\zeta)^{2 j}$. Similar properties hold
for its counterpart related to the derivative with respect to ${\cal B}$:
\begin{eqnarray}
\label{eqd11}
& & \tilde{D}_{l_1} = \frac{\bar\zeta^{l_1}}{(2 j+ \zeta \bar\zeta)^{2 j}}
\sum_{m_2=0}^{2 j} \tilde{\epsilon}_{l_1,m_2} (\zeta \bar\zeta)^{m_2} \quad , \nonumber\\ \quad 
\tilde{\epsilon}_{l_1,m_2}&=&  \frac{(2 j)! (2 j)^{2 j-m_2}  }{\Gamma(m_2+1) \Gamma(2 j-m_2+1)} 
\left[ \frac{1}{\Gamma(l_1+1)}  (1 - \Theta(l_1-2 j)) \right. \nonumber\\
&& \left. + (-1)^{m_2+l_1}
\sum_{m_1}  \frac{(-1)^{-m_1}}{\Gamma(m_2+l_1-m_1+1) \Gamma(m_1-m_2+1)}
\right] \quad .
\end{eqnarray}
The sum on $m_1$ is also restricted:
\begin{equation}
\label{eqd12}
1 \leq m_2 + l_1 - m_1 \leq l_1 \quad .
\end{equation}
Writing the two quantities with the same indices, one finds 
$\tilde{\epsilon}_{k,m}=\tilde{\gamma}_{k,m} $  so that
\begin{eqnarray}
\label{ajout1}
\frac{\tilde{C_k}}{\zeta^k} = \frac{\tilde{D_k}}{\bar\zeta^k} \quad .
\end{eqnarray}
We will continue to write $\tilde{\epsilon}_{k,m}$ and $\tilde{\gamma}_{k,m} $ differently because
it helps in keeping track of the terms origins.

The last set of coefficients, depending on two indices, have a slightly more complex structure;
each can be written as a sum of two terms. The first is the product of the complex variable
$\zeta$ to a power equal to the difference of indices $k-l$ by a function having the same form
than the ones found for the previous coefficients. The second term involves the opposite power
$l-k$ of the conjugate variable $\bar\zeta$:
\begin{eqnarray}
\label{eqd13}
\tilde{M}_{k_1,l_1} &=& \frac{1}{(2 j+ \zeta \bar\zeta)^{2 j}} \left[ 
\zeta^{k_1-l_1} \sum_{m=0}^{2 j}  \tilde{\mu}_{k_1,l_1,m}^{(1)} (\zeta \bar\zeta)^m
+ \bar\zeta^{l_1-k_1} \sum_{m=0}^{2 j+k_1}  \tilde{\mu}_{k_1,l_1,m}^{(2)} (\zeta \bar\zeta)^m 
\right] \, . 
\end{eqnarray}
The constants $\tilde \mu$ are given by the following expressions:
\begin{eqnarray}
\label{eqd14}
\tilde{\mu}_{k_1,l_1,m}^{(1)} &=&  \frac{1}{k_1!l_1!}  \frac{I_{m,k_1+m-l_1}}{R_{j-k_1-m}^2} \nonumber\\
&+& \sum_{m_2}  \frac{(-1)^{k_1+m-l_1-m_2}}{l_1! \Gamma{(k_1+m-l_1-m_2+1)} \Gamma{(-m+l_1+m_2+1)}}
\frac{I_{m,m_2}}{R_{j-l_1-m_2}^2} \quad , \nonumber\\
\tilde{\mu}_{k_1,l_1,m}^{(2)} &=& (1- \Theta(2 j-m)) \nonumber\\
& & \sum_{m_1=0}^{2 j}  
\frac{(-1)^{l_1+m-k_1-m_1}}{k_1! \Gamma{(l_1+m-k_1-m_1+1)} \Gamma{(-m+k_1+m_1+1)}}
\frac{I_{m_1,m}}{R_{j-k_1-m_1}^2} \nonumber\\
&+ & \sum_{m_1,m_2}  
\frac{(-1)^{2 m-m_2+l_1-k_1-m_1}}{\Gamma{(m-m_2+1)} \Gamma{(l_1-k_1+m-m_1+1)} \Gamma{(k_1-m+m_2+1)} \Gamma{(k_1-m+m_1+1)}}
\nonumber\\
& & \frac{I_{m_1,m_2}}{R_{j-k_1+m-m_2-m_1}} \quad ;
\end{eqnarray}
restrictions similar to Eq.(\ref{eqd10},\ref{eqd12}) also exist here.

We have thus obtained a differential expression of an extension of our star product. As we emphasized
earlier, there is more than one extension. This is encoded in the coefficients
$R_m$; in Eq.(\ref{eqb4}) these numbers are defined only for values of $m$ located in a specific
interval while this is clearly not the case in the last formulas. The extension we 
consider here keeps the same expressions for the $R_m$ with $m$ out of  the interval $[-j,j]$, simply
using the extension of factorials by gamma functions. The formula we have derived so far has 
derivatives of any order. In the next section, we will show how this changes. 

We are now in a position to compute some star products explicitly in the case $j \geq 2$:
\begin{eqnarray}
\label{eqd19}
\zeta \star \zeta &=& \frac{1}{( 2 j + \bar\zeta \zeta)^{2 j}} 
\sum_{k=1}^{2 j}
\sigma_k^{(1)} \zeta^{2 j-k+2} {\bar\zeta}^{2 j-k} \quad , \quad
\bar\zeta \star \bar\zeta = \frac{1}{( 2 j + \bar\zeta \zeta)^{2 j}} \sum_{k=1}^{2 j}
\sigma_k^{(2)} \zeta^{2 j-k} {\bar\zeta}^{2 j-k+2} \, , \,
 \nonumber\\
\zeta \star \bar\zeta &=& \frac{1}{( 2 j + \bar\zeta \zeta)^{2 j}} \sum_{k=0}^{2 j}
\sigma_k^{(3)} \zeta^{2 j-k} {\bar\zeta}^{2 j-k} \quad , \nonumber\\
\bar\zeta \star \zeta &=& \frac{1}{( 2 j + \bar\zeta \zeta)^{2 j}} \sum_{k=0}^{2 j}
\sigma_k^{(4)} \zeta^{2 j-k+1} {\bar\zeta}^{2 j-k+1} \, , \,
\end{eqnarray}
where the  $\sigma_k^{(l)}$ are  constants.

Let us now look more closely at the measure.
Introducing the polar coordinates $(r,\theta)$ linked to the Cartesian system
$(x,y)$ defined by $\eta,\bar\eta= x \pm i y$ , the non 
diagonal part of Eq.(\ref{eqd1}) is satisfied if the measure is rotationally
invariant:
\begin{equation}
\label{eqd22}
 d\eta \, d\bar\eta \, h(\eta,\bar\eta) = 2  dr \, d\theta \, r 
 \, h(r) \quad .
\end{equation}
The diagonal part of Eq.(\ref{eqd1}) reads
\begin{equation}
\label{eqd23}
 \int dr f(r) \frac{r^{1+2 j-2 n}}{ (2 j+r^2)^{2 j}} = 
 \frac{1}{4 \pi} \frac{1}{R_n^2} \quad , \quad
 {where} \quad h(r) = i f(r)  \quad .
\end{equation}
Introducing the function 
\begin{equation}
\label{eqd24}
 \Phi(r) = f(r) \frac{r^{1+2 j}}{ (2 j+r^2)^{2 j}} 
\end{equation}
and the variable $s = 1 - 2 \, n$, one obtains the integral equation
\begin{equation}
\label{eqd25}
\int_0^\infty dr \, \Phi(r) \, r^{s-1} = 
\frac{1}{4 \pi} R^2\left(\frac{1}{2} (1-s)\right) \quad .
\end{equation}
This is solved by the inverse Mellin  transformation, leading to
\begin{equation}
\label{eqd26}
h(r) = \frac{1}{8  \pi^2} \frac{(2 j+r^2)^{2 j} }{r^{1+2 j} }  
\int_{-i\infty}^{+i\infty} ds \, \frac{r^{-s}}{R^2\left(\frac{1}{2} 
(1-s)\right)} \quad .
\end{equation}
The quantity $R_n$ has been promoted to a function $R(n)$ by analytical
continuation.

As already pointed out, the measure is not unique. The formula given in 
Eq.(\ref{eqd1}) has to
be satisfied  only for integers $n$ which absolute values are smaller than $j$;
the solution we gave above is obtained by
enforcing it for all real numbers. Till now our treatment is  very similar to \cite{pinzul1} so that the behavior for large $j$
is a priori the same.

\subsection{Restriction on the primary functional space.}

The primary definition of the star product relies on functions defined trough the 
mean values of  operators of the non commuting variables. When $j$ is fixed, the operator
 $ \hat{A} = \sum_{i_1,i_2,i_3=0}^\infty A_{i_1,i_2,i_3} \hat{x}_1^{i_1} 
\hat{x}_2^{i_2} 
\hat{x}_3^{i_3} $ 
can be mapped to a $(2 j+1) \times (2 j+1)$ matrix.
It is thus more 
economical to parameterize it by its entries. In fact, each
function of the form given by Eq.(\ref{eqc1}) can be written as
\begin{equation}
\label{eqe1}
{\cal A}(\zeta,\bar\zeta) = \left[ \frac{1}{(2 j+ \zeta \bar\zeta)^{2 j}} \sum_{m,n=-j}^j 
\tilde{a}_{m,n} R_m R_n 
\bar{\zeta}^{-m+j} \zeta^{-n+j} \right]   \quad {\rm with} \quad 
\tilde{a}_{m,n}= \langle m \vert \hat A \vert n \rangle  \quad .
\end{equation}
For simplicity, we shall rather use the quantities $a_{m,n} = \tilde{a}_{m,n} R_m R_n$.

We now wish to study how the star product we have computed reads when restricted to the primary 
functional 
space whose elements ${\cal A}$ are given in Eq.(\ref{eqe1}). These functions being rational expressions
with numerators and denominators of degree $2 j$ in $\zeta$, one has
\begin{eqnarray}
\label{eqe2}
\frac{\partial^{2 j+1}  }{\partial \zeta^{2 j+1}} 
\left( (2 j + \zeta \bar\zeta)^{2 j} {\cal A}(\zeta,\bar\zeta)  \right) = 
\frac{\partial^{2 j+1}  }{\partial \zeta^{2 j+1}} \left( \sum_{m,n=0}^{2 j}
a_{j-m,j-n} \zeta^m \bar\zeta^n \right) = 0  
\end{eqnarray}
so that the derivative of order $2 j+1$ of such a function can be written as a linear 
combination of those of lower order.

We can now write any derivative of order higher than $2 j$ in terms of those of order
lower than or equal to $2 j$:
\begin{eqnarray}
\label{eqe4}
\frac{\partial^{2 j+k}  }{\partial \zeta^{2 j+k}} {\cal A}(\zeta,\bar\zeta)  &=&
\sum_{m=1}^{2 j} P_{2 j+k,m}  \frac{\partial^m }{\partial \zeta^m}  {\cal A}(\zeta,\bar\zeta)
\quad .
\end{eqnarray}
 One similarly defines a set
of functions $Q_{k,l}$ linked to the derivatives with respect to the conjugate variable $\bar\zeta$:
\begin{eqnarray}
\label{ajout2}
\frac{\partial^{2 j+k}  }{\partial \bar\zeta^{2 j+k}} {\cal A}(\zeta,\bar\zeta)  &=&
\sum_{m=1}^{2 j} Q_{2 j+k,m}  \frac{\partial^m }{\partial \bar\zeta^m}  {\cal A}(\zeta,\bar\zeta)
\quad .
\end{eqnarray}
This allows us to write
the star product with a finite number of derivatives on the primary functional space.

One thus ends up with the following expression of the star product:
\begin{eqnarray}
\label{eqe7}
& & {\cal A}(\zeta,\bar\zeta) \star {\cal B}(\zeta,\bar\zeta)
=  {\cal A}(\zeta,\bar\zeta) \, {\cal B}(\zeta,\bar\zeta) +
{\cal A}(\zeta,\bar\zeta) \, 
\left( \sum_{m=1}^{2 j} 
D_{m} \partial_{\bar\zeta}^{m}  {\cal B}(\zeta,\bar\zeta)
\right)  \nonumber\\
&+& \left(  \sum_{m=1}^{2 j} 
C_{m} \partial_{\zeta}^{m} {\cal A}(\zeta,\bar\zeta) \right)  
 {\cal B}(\zeta,\bar\zeta)  
 +  \sum_{k_1=1}^{2 j}  \sum_{l_1=1}^{2 j}   
M_{k_1,l_1} \partial_{\zeta}^{k_1} {\cal A}(\zeta,\bar\zeta) 
 \partial_{\bar\zeta}^{l_1} {\cal B}(\zeta,\bar\zeta)  \quad .
\end{eqnarray}
The remaining coefficients also involve the functions $P_{k,l}$ and $Q_{k,l}$ :
\begin{eqnarray}
\label{eqe8}
D_m &=&   \left( \tilde{D}_m + \sum_{k=1}^{\infty}  \tilde{D}_{2 j+k} 
 Q_{2 j+k,m} \right)   \quad ,    \nonumber\\ 
M_{k_1,k_2} &=& \tilde{M}_{k_1,k_2} +
\sum_{k=1}^{\infty} \tilde{M}_{k_1,2 j+k} \,  Q_{2 j+k,k_2}  \,
+ \sum_{k=1}^{\infty} \tilde{M}_{2 j+k,k_2} \,  P_{2 j+k,k_1} \nonumber\\
&+& \sum_{k,l=1}^{\infty} \tilde{M}_{2 j+k,2 j+l} \,  P_{2 j+k,k_1} \, Q_{2 j+l,k_2}  \quad .   
 \end{eqnarray}

We have thus shown that although the integral form of the star product contains an infinite
number of derivatives, the 
situation is radically different once we look at it from the primary functional space
perspective. The star product we end up with shares an important property with the one
derived by H.Grosse and P.Presnajder  in the sense that the highest derivatives are of order
$2 j$.  

To go ahead and clarify the structure of the new coefficient functions, we need to write down 
explicitly the dependence of the auxiliary functions $P_{k,l}$ and $Q_{k,l}$.
One easily shows that they are all rational expressions:
\begin{eqnarray}
\label{eqe9}
P_{2 j+k,m} &=& p_{2 j+k,m}  \bar\zeta^{2 j+k-m} (2 j+ \zeta \bar\zeta)^{-(2 j+k-m)} \quad , \nonumber\\
Q_{2 j+k,m} &=& q_{2 j+k,m}  \zeta^{2 j+k-m} (2 j+ \zeta \bar\zeta)^{-(2 j+k-m)}
\quad .
\end{eqnarray}

We are now ready to write the first coefficient:
\begin{eqnarray}
\label{eqe11}
C_m &=& \frac{\zeta^m}{(2 j+\zeta \bar\zeta)^{2 j}} \nonumber\\
& & \sum_{m_1=0}^{2 j} (\zeta \bar\zeta)^{m_1}  \left[ \tilde{\gamma}_{m,m_1} 
  + \sum_{k=1}^{\infty}  p_{2 j+k,m} \, 
\tilde{\gamma}_{2 j+k,m_1} \,  
\left( \frac{\zeta\bar\zeta}{2 j+ \zeta\bar\zeta} \right)^{2 j+k-m} \right]  \quad .
\end{eqnarray}
Compared to the one derived in the previous section, it similarly  has  a power(actually the same) of the complex
variable $\zeta$ multiplied by a function of  $\zeta \bar\zeta$. There is
 an important difference which is that the function involved is a priori not rational
anymore, unless a conspiracy takes place between the quantities $\tilde{\gamma}_{m,n} \quad , \quad
 p_{2j+k,m}$
and $q_{2 j+k,m}$. Restricting ourselves to the primary functional space, we obtain that the star
product gets simpler and more complicated at the same time. It gets simpler in the sense that it
it contains a finite number of terms but more complicated because the coefficient functions $C_k,D_k$ and $M_{k,l}$
do not look rational any more but seem to be replaced by series. The second family of coefficients has a
similar form:
\begin{eqnarray}
\label{eqe12}
D_m &=& \frac{\bar\zeta^m}{(2 j+\zeta \bar\zeta)^{2 j}} \nonumber\\
& & \sum_{m_1=0}^{2 j} (\zeta \bar\zeta)^{m_1}  \left[ \tilde{\epsilon}_{m,m_1} 
  + \sum_{k=1}^{\infty}  q_{2 j+k,m} \, 
\tilde{\epsilon}_{2 j+k,m_1} \,  
\left( \frac{\zeta\bar\zeta}{2 j+ \zeta\bar\zeta} \right)^{2 j+k-m} \right]  \quad .
\end{eqnarray}
The  coefficients with two indices face a similar situation but in addition there is a mixing between
the powers of the two variables which was absent in the extension obtained above.
This is recorded  in the third term of the formula:
\begin{eqnarray}
\label{eqe13}
M_{k_1,k_2} &=& \zeta^{k_1-k_2}
\mu_{k_1,k_2}^{(1)}(\zeta \bar\zeta) + 
\bar\zeta^{k_2-k_1} 
\mu_{k_1,k_2}^{(2)}(\zeta \bar\zeta) +  
\zeta^{-k_2} \bar\zeta^{-k_1} \mu_{k_1,k_2}^{(3)}(\zeta \bar\zeta) 
\end{eqnarray}
where
\begin{eqnarray}
\label{eqe14}
& & \mu_{k_1,k_2}^{(1)}(\zeta \bar\zeta) = \frac{1}{(2 j+\zeta \bar\zeta)^{2 j}}\sum_{m=0}^{2 j} \left[ \tilde{\mu}_{k_1,k_2,m}^{(1)} + \sum_{k=1}^{\infty}
q_{2 j+k,k_2} \, \tilde{\mu}_{k_1,2 j+k,m}^{(1)} \, 
\frac{1}{(2 j+ \zeta \bar\zeta)^{2 j+k-k_2}} \right] (\zeta \bar\zeta)^m \quad , 
\end{eqnarray}
\begin{eqnarray}
& & \mu_{k_1,k_2}^{(2)}(\zeta \bar\zeta) = \frac{1}{(2 j+\zeta \bar\zeta)^{2 j}} 
\sum_{m=0}^{2 j} \left[ \tilde{\mu}_{k_1,k_2,m}^{(2)}  \right. \nonumber\\ 
&& \left. + \sum_{k=1}^{\infty}
p_{2 j+k,k_1} \, \tilde{\mu}_{2 j+k, k_2,m}^{(2)} \, 
\frac{1}{(2 j+ \zeta \bar\zeta)^{2 j+k-k_1}} (1 - \Theta(2 j-m) ) \right] (\zeta \bar\zeta)^m 
\quad {\rm and} \nonumber\\
\end{eqnarray}
\begin{eqnarray}
& & \mu_{k_1,k_2}^{(3)}(\zeta \bar\zeta) =  
\frac{(\zeta \bar\zeta)^{2 j} }{(2 j+ \zeta \bar\zeta)^{4 j}}   
 \sum_{m=0}^{2 j} \left[
 \sum_{k=1}^{\infty} \frac{(\zeta \bar\zeta)^{k}}{(2 j+\zeta \bar\zeta)^{k}}
\left( \tilde{\mu}_{2 j+k,k_2,m}^{(1)} \, p_{2 j+k,k_1} \,
(2 j+\zeta \bar\zeta)^{k_1} \right. \right. \nonumber\\
&& \left. \left. + \tilde{\mu}_{k_1,2 j+k,m}^{(2)} \, q_{2 j+k,k_2} \,
(2 j+\zeta \bar\zeta)^{k_2} \right) \right]  
 (\zeta \bar\zeta)^{m}  \nonumber\\ 
&+& \frac{1}{(2 j+ \zeta \bar\zeta)^{6 j}} \sum_{m=0}^{2 j} \sum_{k,l=1}^{\infty} \frac{p_{2 j+k,k_1,m} \, 
q_{2 j+l,k_2,m}}{(2 j+ \zeta \bar\zeta)^{k+l-k_1-k_2}}
\left[ (\zeta \bar\zeta)^{2 j+k} \sum_{m=0}^{2 j} \tilde{\mu}_{2 j+k,2 j+l,m}^{(1)} 
  \right. \nonumber\\
&& \left. + (\zeta \bar\zeta)^{2 j+l} \sum_{m=0}^{2 j} \tilde{\mu}_{2 j+k,2 j+l,m}^{(2)} 
\right] (\zeta \bar\zeta)^{m} \quad .
\end{eqnarray}

To summarize, the restriction on the primary space is given by 
Eqs.(\ref{eqe7},\ref{eqe11},\ref{eqe12},\ref{eqe13},\ref{eqe14}).
The coefficients obtained so far satisfy the  relation
\begin{eqnarray}
\label{eqe15}
M_{k,l}(\zeta,\bar\zeta) = M_{l,k}(\bar\zeta,\zeta) 
\end{eqnarray}
and the diagonal elements $M_{k,k}$ are   functions  of $ \zeta \bar\zeta $
only.

\section{Link to the H.Grosse and P.Presnajder star product.}

In the last section we showed that the star product derived from our generalized squeezed states
has a finite number of derivatives on its primary functional space. We also saw that the highest  derivative was
of order $2 j$. These two characteristics are shared by the Grosse-Presnajder star product which is expressed in terms of three coordinates:
\begin{eqnarray}
\label{eqd15}
& & f(x_1,x_2,x_3) \star g(x_1,x_2,x_3) = f(x_1,x_2,x_3)  g(x_1,x_2,x_3) \nonumber\\
&+& \sum_{m=1}^{2 j} \frac{(2 j-m)!}{m!(2 j)!} J^{a_1 b_1} J^{a_2 b_2}
\cdots J^{a_m b_m} \nonumber\\
& & \partial_{a_1 a_2 \cdots a_m}  f(x_1,x_2,x_3) \,
\partial_{b_1 b_2 \cdots b_m}  g(x_1,x_2,x_3) \quad .
\end{eqnarray}
The three coordinates are dependent; they  verify the relation $x_1^2+x_2^2+x_3^2 \equiv x^2=1$.
The tensor
\begin{eqnarray}
\label{eqd16}
J^{a b} = x^2  \delta^{a b}  - x^{a} x^{b} + i \epsilon^{a b c} x_c  
\end{eqnarray}
is polynomial, contrary to our coefficient functions which are series of rational expressions. This product
contains mixed derivatives like 
\begin{eqnarray}
\label{eqd17}
J^{1 2} J^{1 2}
\partial_{x_1} \partial_{x_2} f(x_1,x_2,x_3) \partial_{x_1} \partial_{x_2} g(x_1,x_2,x_3) \quad , 
\end{eqnarray}
contrary to the one obtained here for which terms such  as
\begin{eqnarray}
\label{eqd18} 
\partial_{\zeta} \partial_{\bar\zeta} f(\zeta, \bar\zeta) 
\partial_{\zeta} \partial_{\bar\zeta} f(\zeta, \bar\zeta) 
\end{eqnarray}
do not appear. Another  difference is that our formula contains terms such as 
${\cal A} \partial_{\zeta} {\cal B}$ where one of the two functions is not acted upon by a derivative;
nothing of the sort happens in Eq.(\ref{eqd15}).
Despite these apparent differences, we will now show that  there is actually a profound link between 
the two star products.

The star product between two 
elements of the primary functional space is  given by Eq.(\ref{eqc2}). It takes, after 
re parameterization($\zeta=\sqrt{2 j} z$), the form
\begin{eqnarray}
\label{eqf9}
& & \left[ \frac{1}{(1+ z \bar z)^{2 j}} \sum_{m,n=0}^{2 j}
a_{m-j,n-j}
\bar{z}^{2 j-m} z^{2 j-n} \right]
 \star
\left[ \frac{1}{(1+ z \bar z)^{2 j}} \sum_{m,n=0}^{2 j}
b_{m-j,n-j}
\bar{z}^{2 j-m} z^{2 j-n} \right]
 \nonumber\\
&=& \left[ \frac{1}{(1+ z \bar z)^{2 j}} \sum_{m,n=0}^{2 j}
\left( \sum_{l=0}^{2 j} \frac{(2 j-l)!l!}{(2 j)!} a_{m-j,l-j} \, b_{l-j,n-j}  \right)
\bar{z}^{2 j-m} z^{2 j-n} \right] \quad .
\end{eqnarray}

Let us begin with the simplest possibility, $j=1/2$. The generalized squeezed states read
\cite{lubo1}
\begin{eqnarray}
\label{eqg2}
\vert z \rangle = \frac{1}{\sqrt{1 + z \bar z }}
\left( z \vert \frac{1}{2} \rangle + \vert - \frac{1}{2} \rangle \right) \quad .
\end{eqnarray}
An operator built on the algebra generated by the operators 
$\hat{x}_k$
can  be written as a
two by two matrix whose entries will be denoted $a_{k,l}$. The associated
function reads
\begin{equation}
\label{eqg3}
{\cal A} (z,\bar z) = \frac{a_{-\frac{1}{2},-\frac{1}{2}} z \bar{z} + a_{\frac{1}{2},-\frac{1}{2}} z +
a_{-\frac{1}{2},\frac{1}{2}} \bar z + a_{\frac{1}{2},\frac{1}{2}}}{1+ z \bar z}\quad .
\end{equation}  
Let us consider another operator , whose entries are labeled
$b_{k,l}$ and which generate the function ${\cal B}$. We obtain for the star 
product of 
such functions, according to Eq.(\ref{eqf9}).
To obtain the canonical expression of the star product, one needs to subtract the
ordinary product
of the  two functions:
\begin{eqnarray}
\label{eqg5}
& & {\cal A} (z,\bar z) \star {\cal B} (z,\bar z)  -
 {\cal A} (z,\bar z)  {\cal B} (z,\bar z) \nonumber\\
&=& \frac{1}{(1 + z \bar{z})^2} \left( - a_{\frac{1}{2},-\frac{1}{2}} + 
 \left( a_{\frac{1}{2},\frac{1}{2} } - a_{-\frac{1}{2},-\frac{1}{2}} \right) \bar{z} + 
 a_{-\frac{1}{2},\frac{1}{2}} \bar{z}^2  \right) \nonumber\\
& & \left( - b_{-\frac{1}{2},\frac{1}{2}} + 
 \left( b_{\frac{1}{2},\frac{1}{2} } - b_{-\frac{1}{2},-\frac{1}{2}} \right) z + 
 b_{\frac{1}{2},-\frac{1}{2} } z^2  \right)
   \quad .
\end{eqnarray}

We need to write the right member of this equation in terms of derivatives. For that,
we have to find the expressions of $a_{-\frac{1}{2},-\frac{1}{2}}, \cdots $ in terms of the derivatives 
of the function
${\cal A}$ and  $b_{-\frac{1}{2},-\frac{1}{2}}, \cdots$ in terms of those of the function ${\cal B}$.
Inspired by what was obtained in the previous section, we shall use derivatives of order
$1$ or lower in each of the two variables. The result reads
\begin{eqnarray}
\label{eqg6}
a_{\frac{1}{2},\frac{1}{2}} &=&  (1 - \bar z \partial_{\bar z} +
z ( -\partial_z + \bar z (1 + z \bar z ) \partial_z \partial_{\bar z} ) ) {\cal A}(z,\bar z)  \quad , \nonumber\\
a_{-\frac{1}{2},-\frac{1}{2}} &=&  (1 + \bar z \partial_{\bar z} +
z ( \partial_z + \bar z (1 + z \bar z ) \partial_z \partial_{\bar z} ) ) {\cal A}(z,\bar z) \quad , \nonumber\\
a_{-\frac{1}{2},\frac{1}{2}} &=& (1 
- z (z \partial_z + (1 + z \bar z) \partial_z \partial_{\bar z} )
 {\cal A}(z,\bar z)  \quad , \nonumber\\
a_{\frac{1}{2},-\frac{1}{2}} &=& (- \bar{z}^2 \partial_{\bar z} + \partial_z -
\bar z (1+z \bar z) \partial_z \partial_{\bar z} ) {\cal A}(z,\bar z) \quad .
\end{eqnarray}
After a straightforward computation, one finds  the star product can be written in a very simple 
way:
\begin{equation}
\label{eqg7}
 {\cal A}(z,\bar z) \star  {\cal B}(z,\bar z) =
 {\cal A}(z,\bar z) \,  {\cal B}(z,\bar z) +
(1 + \bar z z)^2  \, \frac{\partial}{\partial z}
{\cal A}(z,\bar z) \, \frac{\partial}{\partial \bar z}
 {\cal B}( z,\bar z) \quad .
\end{equation}

At this point one can point out the non trivial fact that although mixed derivatives with 
respect to $z$ and $\bar z$ appear in the expression of $a_{-\frac{1}{2},-\frac{1}{2}}, \cdots$, the final
formula does not contain such terms. This was of course predicted in the last section,
using the measure and the completeness relation. This is recovered here.

Once the star product is built, one has to verify for example that the functions
corresponding to the basic operators have star commutators which reproduce the
quantum commutation relations. In our case one can effectively verify that
\begin{equation}
\label{eqg8}
[ {\cal X}_k(z , \bar z) ,  {\cal X}_l(z , \bar z) ]_\star =
 {\cal X}_k(z , \bar z) \star  {\cal X}_l(z , \bar z)
 - {\cal X}_l(z , \bar z) \star  {\cal X}_k(z , \bar z) = 
\frac{i}{\sqrt{j(j+1)}} \epsilon_{klr}  {\cal X}_r(z , \bar z) \quad ,
\end{equation}
where
\begin{eqnarray}
\label{ajout4}
{\cal X}_k(z,\bar z) = \langle z \vert \hat{x}_k \vert z \rangle .
\end{eqnarray}
This is easily verified since these functions read \cite{lubo1}
\begin{equation}
\label{eqg9}
{\cal X}_1(z,\bar z) = \frac{1}{\sqrt{3}} \frac{\bar z + z}{1 + \bar z z} \quad , \quad
{\cal X}_2(z,\bar z) = - i \frac{1}{\sqrt{3}} \frac{\bar z - z}{1 + \bar z z} \quad , \quad
{\cal X}_3(z,\bar z) = 
\frac{1}{\sqrt{3}} \frac{\bar z  z - 1}{1 + \bar z z} \quad .
\end{equation}
Concerning  associativity, one readily obtains
\begin{eqnarray}
\label{eqg10}
& & ( f(z,\bar z) \star g(z,\bar z) ) \star h(z,\bar z) -
f(z,\bar z) \star ( g(z,\bar z) \star h(z,\bar z)) \nonumber\\
&&=
- (1+z \bar z)^3
\left( - ( 2 \, z \, \partial_{\bar z} \, h(z,\bar z) +
(1+z \bar z) \, \partial_{\bar z}^2 \, h(z,\bar z))
\partial_{z} \, f(z,\bar z) \, \partial_{z} \, g(z,\bar z) \right. \nonumber\\
&& \left. + ( 2 \, \bar z \, \partial_{z} \, f(z,\bar z) 
+ (1 + z \bar z) \,
\partial_{z}^2 \, f(z,\bar z) ) 
\partial_{\bar z} g(z,\bar z) \partial_{\bar z}
h(z,\bar z)  \right)
\end{eqnarray}
which does not vanish for all functions $f,g,h$ but does so on the quotient of polynomials
displayed in Eq.(\ref{eqe1}). For example, 
$z \star (z \star \bar z) - (z \star z  ) \star \bar z = 
-2 \bar z (1+ z\bar z)^3$ \quad .

We emphasize that the form of the star product given above is
associative only for the set of functions appearing in Eq.(\ref{eqe1}).
All the elements of this set go to a constant at infinity on the complex plane. Therefore, this family
doesn't contain the function $z$ for example.

Let us now come back to the Weyl correspondence. We will also  illustrate this in the case
$j=1/2$. To each function of the form displayed in Eq.(\ref{eqg3}), one associates
an operator given by
\begin{eqnarray}
\label{ajout11}
\hat W [f]  &=& W_1[f] \hat{x}_1 + W_2[f] \hat{x}_2 + W_3[f] \hat{x}_3 + 
W_4[f]  \quad {\rm where} \nonumber\\
W_1[f] &=& \frac{1}{2} (  a_{-\frac{1}{2}, \frac{1}{2}} + 
a_{\frac{1}{2},- \frac{1}{2}}) \quad , \quad
W_2[f] = \frac{i}{2} ( - a_{-\frac{1}{2}, \frac{1}{2}} + 
a_{\frac{1}{2},- \frac{1}{2}}) \quad , \quad \nonumber\\
W_3[f] &=& \frac{1}{2} (  a_{\frac{1}{2}, \frac{1}{2}}
 - a_{-\frac{1}{2},- \frac{1}{2}}) \quad , \quad
W_4[f] = \frac{i}{2} ( a_{\frac{1}{2}, \frac{1}{2}}
 + a_{-\frac{1}{2},- \frac{1}{2}}) \quad .
\end{eqnarray}
One can then see that Eq.(\ref{eqc1}) is verified.
These coefficients can be written in terms of integrals of the function:
\begin{eqnarray}
\label{ajout22}
a_{-\frac{1}{2}, \frac{1}{2}} &=& \frac{3 i}{\pi}  \int dz d \bar z   
\frac{ z}{(1+ z \bar z)^3} f(z,\bar z) \quad , \quad
a_{\frac{1}{2},-\frac{1}{2}} = \frac{3 i}{\pi}  \int dz d \bar z   
\frac{\bar z}{(1+ z \bar z)^3} f(z,\bar z) \quad , \quad \nonumber\\
a_{-\frac{1}{2},-\frac{1}{2}} &=& - \frac{ i}{\pi}  \int dz d \bar z   
\frac{ 1 + 2 z \bar z}{(1- z \bar z)^3} f(z,\bar z) \quad , \quad
a_{\frac{1}{2},\frac{1}{2}} =  \frac{ i}{\pi}  \int dz d \bar z   
\frac{ 2 - z \bar z}{(1+ z \bar z)^3} f(z,\bar z) \quad .
\end{eqnarray}
The formula given in Eq.(\ref{ajout22}) when used in Eq.(\ref{ajout11})  allows one to write
the image of a function by the Weyl map as an integral on this function, as
in \cite{stei}.

To have a broader view, lets us consider the cases $j=1$ and $j=2$. One can similarly to 
Eq.(\ref{eqg6}) obtain the coefficients on the functions ${\cal A}(z,\bar z)$ in terms of its
derivatives, thanks to the fact that the primary functional space is finite dimensional. Replacing
them in Eq.(\ref{eqf9}), one obtains the star product in terms of differential operators. We record the
 results in the variable $z=\sqrt{2 j} z$ in which they look much simpler and will allow
 a direct comparison
with the Grosse-Presnajder's. In both cases, one has
\begin{equation}
\label{eqg11}
{\cal A}(z,\bar z) \star {\cal B}(z,\bar z) =
{\cal A}(z,\bar z)  {\cal B}(z,\bar z) +
\sum_{k,l=1}^{2 j} M_{k,l} \, \partial_{z}^k {\cal A}(z,\bar z)  \,
\partial_{\bar z}^l {\cal B}(z,\bar z)
\end{equation}
where the coefficient functions are polynomials.
In the case $j=1$, one has
\begin{eqnarray}
\label{ajout10}
M_{2,2} &=& \frac{1}{4} (1+z \bar z)^4 \, , \,
M_{2,1} = \frac{1}{2}  z (1+z \bar z)^3  \, , \,
M_{1,1} = \frac{1}{2} (1 + 2 z \bar z) (1+z \bar z)^2  \quad .
\end{eqnarray}
Formulas get more complicated as one increases the value of the cut off; for $j=2$, one obtains
\begin{eqnarray}
\label{eqg12}
M_{4,4} &=& \frac{1}{576} (1+z \bar z)^8 \quad , \quad
M_{4,3} = \frac{1}{48} z (1+z \bar z)^7 \quad , \quad
M_{4,2} = \frac{1}{16} z^2 (1+z \bar z)^6 \quad , \nonumber\\
M_{4,1} &=& \frac{1}{24} z^3 (1+z \bar z)^5 \quad , \nonumber\\ 
M_{3,3} &=& \frac{1}{144} (1+36 z \bar z) (1+z \bar z)^6 \quad , \quad
M_{3,2} = \frac{1}{24} z (1+18 z \bar z) (1+z \bar z)^5 \quad , \nonumber\\
M_{3,1} &=& \frac{1}{24} z^2 (1+ 12 z \bar z) (1+z \bar z)^4 \quad , \nonumber\\
M_{2,2} &=& \frac{1}{24} (1+ 6 z \bar z + 54 z^2 \bar{z}^2) (1+z \bar z)^4 \quad , \quad
M_{2,1} = \frac{1}{12} z (1+ 3 z \bar z + 18 z^2 \bar{z}^2) (1+z \bar z)^3 \quad , \nonumber\\
M_{1,1} &=& \frac{1}{12}  (3+ 2 z \bar z + 3 z^2 \bar{z}^2 + 12 z^3 \bar{z}^3) (1+z \bar z)^2 \quad .
\end{eqnarray}
The coefficient functions verify the symmetry property $M_{k,l}(z,\bar z) = M_{l,k}(\bar z, z)$ 
so that we wrote  only the upper half of the matrix. 

From these examples we will make a few observations, hoping they will prove to be true in the general
case. First, the coefficients $C_k,D_k$
appearing in Eq.(\ref{eqe7}) seem to be vanishing. The second fact is that the coefficient functions are polynomials with a special structure; they contain
specific powers of $(1+ z \bar z)$:
\begin{eqnarray}
\label{eqg14}
M_{k,l} = z_{k,l} \,  N_{k,l} \, (1+ z \bar z)^{k+l} \quad ,
\end{eqnarray}
where the quantities $z_{k,l}$ are constants while  $N_{k,l}$ are polynomials of
degree $4 j-(k+l)$ obeying the following properties:
\begin{itemize}
     \item    The polynomial $ N_{1,1} $ is a function of the product 
              $ z \bar z $  of degree  $2 j - 1$:
	      \begin{equation}
              \label{eqg15}
               N_{1,1} = (1 + a_1 z \bar z   + \cdots
	       + a_{2 j-1} (z \bar z)^{2 j-1} ) \quad .
	       \end{equation}
    \item    The others are  derivatives of the previous one:
              \begin{equation}
              \label{eqg16}
	      N_{k,l}  = \partial_{\bar z}^{k-1} \partial_{z}^{l-1} N_{1,1} \quad .
	      \end{equation}
\end{itemize}

Although the star product we obtained here may look superficially "new", it seems in fact to be
nothing else than the one obtained by Grosse-Presnajder, but  written in stereographic coordinates.
We hereafter show that this is what happens for the lower values of $j$. 

One may write the expression of the star product displayed in Eq.(\ref{eqd15}) in stereographic 
coordinates, replacing partial derivatives by covariant derivatives. We found simpler a mixed 
presentation which we now present. From the definition of the stereographic coordinates
\begin{eqnarray}
\label{eqf2}
x_1 = \frac{\bar z + z}{1+ z \bar z} \quad , \quad
x_2 = - i  \frac{\bar z - z}{1+ z \bar z} \quad , \quad
x_3 = \frac{z \bar z  - 1}{1+z \bar z} \quad ,
\end{eqnarray}
one obtains that the tensor $J^{a b}$ becomes  the following
{\sl non symmetric} matrix:
\begin{eqnarray}
\label{eqg17}
J^{1,1} &=& \frac{ (-1+{\bar z}^2) (-1+ z^2) }{(1+ \bar z z)^2} \, , \,
J^{1,2} = i \frac{ (-1+{\bar z}^2) (1+ z^2) }{(1+ \bar z z)^2} \, , \,
J^{1,3} = - \frac{2  (-1+{\bar z}^2 z) (-1+ z^2) }{(1+ \bar z z)^2} \, , \nonumber\\
J^{2,1} &=& - i \frac{ (1+{\bar z}^2) (-1+ z^2) }{(1+ \bar z z)^2} \, , \,
J^{2,2} =  \frac{ (1+{\bar z}^2) (1+ z^2) }{(1+ \bar z z)^2} \, , \,
J^{2,3} = \frac{2 i (1+{\bar z}^2 ) z }{(1+ \bar z z)^2} \, , \nonumber\\
J^{3,1} &=& - \frac{2 \bar z (-1+ z^2 )  }{(1+ \bar z z)^2} \, , \,
J^{3,2} = - \frac{2 i \bar z (1+{\bar z}^2 )  }{(1+ \bar z z)^2} \, , \,
J^{3,3} = \frac{ 4 \bar z z }{(1+ \bar z z)^2} \quad .
\end{eqnarray}
The link between the derivatives with respect to the two set of coordinates
\begin{eqnarray}
\label{eqg18}
\partial_{x_1} f &=& \frac{1}{2} (1+ z \bar z) ( \partial_{ z} f +
\partial_{\bar z} f )
\quad , \quad
\partial_{x_2} f = \frac{i}{2} (1+ z \bar z) (- \partial_{ z} f +
\partial_{\bar z} f ) \quad , \nonumber\\
\partial_{x_3} f &=& \frac{1}{2} (1+ z \bar z) (z \partial_{ z} f +
\bar z \partial_{\bar z} f ) \quad ,
\end{eqnarray}
will be summarized by the following formula
\begin{eqnarray}
\label{eqg19}
\partial_{x_a} f & \equiv & A_a \partial_{ z} f + B_a
\partial_{\bar z} f  \quad .
\end{eqnarray}
It is straightforward to verify  some relations  between the tensor $J$ and the vectors
$A,B$:
\begin{eqnarray}
\label{eqg20}
J^{a,b} A_a A_b =  J^{a,b} B_a A_b =
J^{a,b} B_a B_b = 0 \quad , \quad
J^{a,b} A_a B_b = (1 + z \bar z)^2 \quad .
\end{eqnarray}
Let us consider the case $j=1/2$. From the previous relations, one obtains
\begin{eqnarray}
\label{eqg21}
& & f(z,\bar z) \star g(z,\bar z) -
f(z,\bar z)  g(z,\bar z) \nonumber\\
&=&   J^{a,b} ( A_a \partial_{ z} f(z,\bar z) + B_a
\partial_{\bar z} f(z,\bar z) ) ( A_b \partial_{ z} g(z,\bar z) + B_b
\partial_{\bar z} g(z,\bar z) ) \nonumber\\
&=& J^{a,b} A_a B_b  \partial_{ z} f(z,\bar z)
\partial_{ \bar z} g(z,\bar z) = (1 + z \bar z)^2  \partial_{ z} f(z,\bar z)
\partial_{ \bar z} g(z,\bar z)
\end{eqnarray}
which is exactly the expression we obtained with our star product in
Eq.(\ref{eqd9}).

We have verified that doing the same for $j=1$ and $j=2$, one obtains exactly the results obtained
in Eq.(\ref{eqg12},\ref{ajout10}).

\section{Spherical harmonics.}

On one side, spherical harmonics are natural functions for systems or theories with spherical 
symmetry. On the other side, the functions we have worked with so far are rational expressions of the complex variable
$z$ and its complex conjugate $\bar z$. Can we make a connection between the two sets? The answer seems
to be positive, provided one uses the stereographic projection. 

When expressed in spherical coordinates, the spherical harmonics read
\begin{eqnarray}
\label{eqf1}
Y_{j,m}(\theta, \phi) = P_j^m(\cos{\theta}) \, e^{i m \phi} \quad .
\end{eqnarray}
where the $P_j^m$ are Legendre functions which turn out to be polynomials for special values of
$j$ and $m$. One straightforwardly  obtains that in the stereographic coordinates,
\begin{eqnarray}
\label{eqf3}
Y_{j,m}(\theta, \phi) =  
\left(\frac{\bar z}{ \sqrt{z \bar z}} \right)^m
P_j^m \left( \frac{z \bar z  - 1}{1+z \bar z} \right) \quad .
 \end{eqnarray}
The Legendre functions have the following form
\begin{eqnarray}
\label{eqf4}    
 \sum_{k,l} p^m_{j,k,l} \left( \frac{z \bar z  - 1}{1+z \bar z} \right)^k
\left(2 \frac{\sqrt{z \bar z} }{1+z \bar z} \right)^l \quad ,
\end{eqnarray}  
the $p^m_{j,k,l}$ being real numbers.
From here one sees that the components of the spherical harmonics will be rational 
like the functions of our primary functional space only
for even values of $l$. For example, in the case $j=1/2$, this does not happen:
\begin{eqnarray}
\label{eqf5}
Y_{\frac{1}{2},-\frac{1}{2}} = \frac{1}{\sqrt{\bar z}} \frac{1-z \bar z}{\sqrt{1+z \bar z}}
\quad , \quad
Y_{\frac{1}{2},\frac{1}{2}} =  \frac{\sqrt{\bar z}}{ \sqrt{1 + z \bar z}} \quad .
\end{eqnarray} 
On the contrary, when $j=1$, this is the case:
\begin{eqnarray}
\label{eqf6}
Y_{1,-1} =   \sqrt{\frac{3}{2 \pi}}   \frac{z}{1+ z \bar z} \quad , \quad
Y_{1,0} =  \frac{1}{2}  \sqrt{\frac{3}{\pi}}   \frac{-1+ z \bar z}{1+ z \bar z} \quad , \quad
Y_{1,1} = - \sqrt{\frac{3}{2 \pi}}   \frac{\bar z}{1+ z \bar z} \quad .
\end{eqnarray}
On the other side, a typical function $\cal{A}$ of the primary functional space takes the form
displayed in Eq.(\ref{eqf9}): it is the quotient of two  polynomials of degree four.
The hypergeometrics are recovered for special choices of the constants $a_{m,n}$: 
\begin{eqnarray}
\label{eqf8}
a_{1,0} &=& a_{0,-1} = \sqrt{ \frac{3}{2 \pi}}   \rightarrow
Y_{1,-1} \quad , \quad
a_{1,1} = - a_{-1,-1} = - \frac{1}{2}  \sqrt{\frac{3}{\pi}}  \rightarrow
Y_{1,0} \quad , \nonumber\\
a_{0,1} &=& a_{-1,0} = - \sqrt{ \frac{3}{2 \pi}}   \rightarrow
Y_{1,1}
\end{eqnarray}
in each case the non mentioned constants are take to be vanishing.

For the specific case at hand, the computation of the star products using the differential 
formulation given  is not the easiest. One readily makes the calculations using Eq.(\ref{eqf9}) and obtains that the star 
products of hypergeometrics 
can be written as sums of linear and  quadratic expressions:
\begin{eqnarray}
\label{eqf10}
Y_{1,-1} \star Y_{1,-1} &=&  \frac{1}{2}  Y_{1,-1}  Y_{1,-1} \quad , \quad
Y_{1,0} \star Y_{1,0} =    Y_{1,0}^2 - Y_{1,1}  Y_{1,-1}  \quad , \nonumber\\
Y_{1,1} \star Y_{1,1} &=&  \frac{1}{2}  Y_{1,1}  Y_{1,1}  \quad , \nonumber\\ 
Y_{1,-1} \star Y_{1,0} &=& Y_{1,-1}  \left( \frac{1}{4} \sqrt{\frac{3}{\pi}} +
\frac{1}{2} Y_{1,0}  \right) \quad , \quad
Y_{1,-1} \star Y_{1,1} = \frac{1}{4}   \left( \sqrt{\frac{3}{\pi}}  Y_{1,0} - 
2 Y_{1,0}^2 + 6 Y_{1,1} Y_{1,-1}  \right)  \quad , \nonumber\\
Y_{1,0} \star Y_{1,-1} &=& - \frac{1}{4} Y_{1,-1}  \left( \sqrt{\frac{3}{\pi}} - 
2 Y_{1,0}  \right) \quad , \quad
Y_{1,0} \star Y_{1,1} =  \frac{1}{4} Y_{1,1}  \left( \sqrt{\frac{3}{\pi}} + 
2 Y_{1,0}  \right) \quad , \nonumber\\
Y_{1,1} \star Y_{1,-1} &=& \frac{1}{4}   \left(- \sqrt{\frac{3}{\pi}}  Y_{1,0} - 
2 Y_{1,0}^2 + 6 Y_{1,1} Y_{1,-1}  \right) \quad , \nonumber\\
Y_{1,1} \star Y_{1,0} &=&  \frac{1}{4} Y_{1,1}  \left( \sqrt{\frac{3}{\pi}} - 
2 Y_{1,0}  \right) \quad .
\end{eqnarray}

The fact that the coefficient functions appearing in the expression of the star product do not contain
complex numbers imply that
\begin{eqnarray}
\label{eqf12}
({\cal A} \star {\cal B})^+ = {\cal B}^+ \star {\cal A}^+
\end{eqnarray}

We have more elements in our primary functional space than there are hypergeometrics
for a fixed value of $j$; this is evident from Eq.(\ref{eqf8}). This is not necessarily a handicap;
when using coherent states, one actually introduces an over complete basis, which amounts to
dealing with a redundant system. For the case we examined in this section, the hypergeometrics
being a subset of our primary space, the expression of the star product we obtained are also
valid for them.

\section{Conclusions.}

A family of generalized squeezed states on the fuzzy sphere has been proposed. It is based
on an uncertainty relation which is weaker than Heisenberg's. In this work, we have constructed the associate
star product. The integral formulation leads to an expression with an infinite number of 
derivatives, with
rational coefficient functions. On the contrary, the restriction to the primary functional 
space is simpler
, like the one obtained by Grosse-Presnajder: it involves derivatives
up to the order $2 j$. The coefficient functions, at first sight, become series rather than being rational
functions. Working out explicitly the cases $j=1/2,1$ and $2$, we found that in fact the coefficient
functions are polynomial: this indicates a conspiracy between the quantities appearing in
Eqs.(\ref{eqe11},\ref{eqe12},\ref{eqe13},\ref{eqe14}), but we did not pursue the analysis in that
direction. The star product has the interesting property that it does not contain mixed 
derivatives: the derivative with respect to the variable $z$ act only on the first argument
while the derivative with respect its conjugate acts only on the second argument. We also explicitly
showed that for the lowest values at least, our star product coincide with the one found by
H.Grosse and P.Presnajder. This provides a physical interpretation of this star product in the sense
that the states on which it is based saturate an uncertainty  which is weaker than the 
Heisenberg relation.

The interest of the formulation presented here is twofold. First,  star products with a finite
number of derivatives play an
important  role in the proof that Q.F.T on the fuzzy sphere are finite. Secondly,
the states used here are parameterized by a complex variable which is linked to the stereographic
coordinates. This opens the possibility of studying field theory by a deformation of the theory
written on the sphere in these coordinates \cite{avenir}, obtaining a kind of  "space-time"
description in contrast to many studies in which it was rather the "angular momentum" representation
which was used. The formulas we have displayed so far display the important features of 
the star product but are not transparent concerning what happens in the large $j$ limit, because of
the intricate dependences of the coefficient functions on $j$. One of the first tasks will be
their simplification or a formulation leading to more tractable formulas from the start.

This work provides
a way to obtain a fuzzy version of the algebra of the compactified  plane.
Unlike standard representations, the maps used are not based on the spherical
harmonics (homogeneous polynomials in $x,y,z$) but on
polynomials in $z$ and $ \bar z$ mapped on a finite dimensional space.


\begin{thebibliography}{99}
\bibitem{prea}
By H. Grosse, P. Presnajder, Lett.Math.Phys.46:61-69,1998.
\bibitem{preb}
By H. Grosse, C. Klimcik, P. Presnajder,
Commun.Math.Phys.180:429-438,1996.
\bibitem{seiberg}  Nathan Seiberg, Edward Witten, JHEP 9909 (1999) 032.
\bibitem{sidha} B.G. Sidharth. physics/0405157
\bibitem{takeh} Takehiro Azuma, Subrata Bal, Keiichi Nagao, Jun Nishimura. 
                hep-th/0405277
\bibitem{castro} P. Castro-Villarreal, R. Delgadillo-Blando, Badis Ydri. 
                 hep-th/0405201.
\bibitem{biga}  Daniela Bigatti, hep-th/0405115.
\bibitem{urs} Ursula Carow-Watamura, Harold Steinacker, Satoshi Watamura,. 
              hep-th/0404130
\bibitem{anyon}  P.A. Horvathy, M.S. Plyushchay, hep-th/0404137.
\bibitem{stei}  Daniel Sternheimer, math.QA/9809056, AIP Conf.Proc. 453 (1998) 107-145.
\bibitem{moyal1} H.Groenewold, Physica(Amsterdam) 12(1946)405.
\bibitem{moyal2} J.Moyal, Proc. Camb. Phil. Soc. 45(1949)99.
\bibitem{voros} A.Voros, Phys. Rev. A40 (1989) 6814.
\bibitem{pre1} H. Grosse, P. Presnajder, Lett.Math.Phys.28:239-250,1993.
\bibitem{pre2} P.Presnajder, J.Math.Phys.41:2789-2804,2000.
\bibitem{wata1} Ursula Carow-Watamura, Satoshi Watamura,
Commun.Math.Phys.183:365-382,1997.
\bibitem{wata2} Ursula Carow-Watamura, Satoshi Watamura,
Int.J.Mod.Phys.A13:3235-3244,1998.
\bibitem{wata3} Ursula Carow-Watamura, Satoshi Watamura,
Commun.Math.Phys.212:395-413,2000.
\bibitem{bala1} A.P. Balachandran, S. Vaidya,Int.J.Mod.Phys.A16:17-40,2001.
\bibitem{bala2} A.P. Balachandran, X. Martin, Denjoe O'Connor,
Int.J.Mod.Phys.A16:2577-2594,2001.
\bibitem{bala3} A.P. Balachandran, T.R. Govindarajan, B. Ydri, hep-th/0006216.
\bibitem{ho} Pei-Ming Ho, JHEP 0012 (2000) 015.
\bibitem{arik} T. Hakioglu, M. Arik, Phys.Rev.D 54(1999)52.
\bibitem{star1}  Keizo Matsubara, Mårten Stenmark. hep-th/0402031.
\bibitem{star2}  K. Hayasaka, R. Nakayama, Y. Takaya,
Phys.Lett. B553 (2003) 109-118. 
\bibitem{star3}  A. P. Balachandran, S. Kurkcuoglu, E. Rojas,
JHEP 0207 (2002) 056.
\bibitem{star4} Brian P. Dolan, Oliver Jahn,
Int.J.Mod.Phys. A18 (2003) 1935-1958
\bibitem{star5} A.B. Hammou, M. Lagraa, M.M. Sheikh-Jabbari,
Phys.Rev. D66 (2002) 025025.
\bibitem{star6} A.P. Balachandran, Brian P. Dolan, J. Lee, X. Martin,
Denjoe O'Connor, J.Geom.Phys. 43 (2002) 184-204.
\bibitem{pinzul1}  G. Alexanian, A. Pinzul, A. Stern,
Nucl.Phys.B600:531-547,2001.
\bibitem{lubo1} Musongela Lubo, JHEP05(2004)052.
\bibitem{madore1} J. Madore, Class.Quant.Grav.9:69-88,1992.
\bibitem{madore2} 
J.Madore, {\sl An Introduction to Noncommutative Differential Geometry
                 and its Applications}, Cambridge University Press,
		 Cambridge, 1995.
\bibitem{pere}  A. Perelomov, {\sl Generalized coherent states and their application},
  Springer (1986).
\bibitem{klau} J.R.Klauder, B.S. Skagerstam, Coherent states(World Scientific,
Singapore,1985).
\bibitem{kon} Maxim Kontsevich, q-alg/9709040. 
\bibitem{za} V.I. Man'ko, G. Marmo, E.C.G. Sudarshan, F. Zaccaria,
 Phys.Scripta 55(1997)528.
\bibitem{nekra} Michael R. Douglas, Nikita A. Nekrasov,
Rev.Mod.Phys.73:977-1029,2001.
\bibitem{avenir} Musongela Lubo, in preparation.
\end{thebibliography}
\end{document}